\documentclass[preview]{elsarticle}

\usepackage{graphicx}
\usepackage{hyperref}
\usepackage{multirow}

\journal{Brazilian Journal of Physics}

\begin{document}

\begin{frontmatter}

\title{Orthogonal and  Non-Orthogonal Tight Binding parameters for III-V Semiconductors Nitrides}
\author[ASM]{A. S. Martins}  
\ead{asmartins@id.uff.br}
\author[CEF]{C. E. Fellows}

\address{Departamento de F\'\i sica - ICEx, Universidade Federal Fluminense}

\begin{abstract}
A simulated annealing (SA) approach is employed in the determination of different tight binding (TB) sets of parameters for the nitride semiconductors AlN, GaN and InN, as well their limitations and potentialities are also discussed. Two kinds of atomic basis set are considered: (\textit{i}) the orthogonal $sp^3s*$ with interaction up to second neighbors and (\textit{ii}) a $spd$ non-orthogonal set, with the Hamiltonian matrix elements calculated within the Extended H\"uckel Theory (EHT) prescriptions. For the non-orthogonal method, TB parameters are given for both zincblend and wurtzite crystalline structures.

\end{abstract}
\begin{keyword}
	A. Semiconductors; A. Insulators; B. Epitaxy; D. Electronic Band Structure
	\PACS 71.20.Eh \sep 71.22.+i \sep 81.05.Ea \sep 71.20.Mq
\end{keyword}

\end{frontmatter}


\section{Introduction}
\label{sec:intro}

Even nowdays, semiconductor systems are widely studied experimentally and theoretically due to their broad application in optoelectronics. Despite the huge technological growth of red and yellow light-emitting diodes (LEDs), there are still efforts to extend their operation into the short wavelength region of the visible spectrum (from green to violet). Although successful attempts to make LEDs and LDs with SiC and II-VI materials (e.g., ZnSe) have been made, the purpose of such devices has been diminished by the very low efficiency in SiC diodes and the short lifetimes of carriers in II-VI materials, which is due to the relative facility of defect formation. As a result, the III-V nitride materials with wurtzite crystal structure (GaN, AlN, InN, and their alloys) have generated considerable interest for operation at these short wavelengths. Since the electronic band structure for each one of the nitride materials possesses a direct transition with a band gap energy range from 1.9 eV (InN) to 6.2 eV for AlN at room temperature as well rather high thermal conductivity, the (AlIn)GaN system has been explored in the areas of high-power and high-temperature electronic devices and short-wavelength (visible and ultraviolet) optoelectronic devices. See \cite{strite1992,vurgaftman2001} for a review of relevant properties of these compounds.

It is known that semiconductor devices and nanostructures simulations demand large supercells, requiring a great computational effort if the systems properties are calculated within first-principle approaches. On the other hand, the theoretical description of large supercells is possible within the semi-empirical tight-binding (TB) formalism with modest computational load. TB calculations employ atomic orbitals as basis set, with the Hamiltonian matrix elements (orbital energies and hoppings) being parameterized in order to reproduce the experimental band structure of the crystalline material. The basis of the TB method applied to Solid State Physics were established in the seminal paper of Slater and Koster \cite{slater1954}, which assumes the atomic orbitals as an orthogonal basis set: the so called Orthogonal Tight Binding (OTB) formalism. The TB parameters in the orthogonal formulation have reduced their transferability for different environments from the ideal crystalline structure they were frist calculated. For instance, the description of pressure effects on the electronic properties of crystals is only possible with a suitable correction of the hopping elements, which are due the atom-atom distance reduction.

One alternative to OTB is to approach the problem within the semi-empirical Extended H\"uckel Theory (EHT)\cite{jcerda2000,asmssc}. As in any TB formalism, the EHT alloy states are spanned as linear combinations of atomic orbitals, considering that the basis orbitals form a non-orthogonal basis set. The aforementioned method presents a good transferability of its parameters \cite{jcerda2000} and it also gives a good description of the deformation potential for III-V alloys under uniform strain \cite{asmssc}. In addition, EHT is capable of reproducing the density of states (DOS) of graphene, silicene and germanene supercells with a single vacancy, in a remarkble agreement with Density Functional Theory (DFT) calculations \cite{asmjpcm}.

The paper is organized as follows. In section \ref{sec:method}, a summary of the theoretical tools employed for studying the alloys is presented, namely the tight binding approach for the electronic structure calculations in both orthogonal and non-orthogonal (H\"uckel) formulation and the Simulated Annealing (SA) method for calculating the TB parameters. Finally, section \ref{sec:results} present all the TB sets in both orthogonal and non-orthogonal formulations and a discussion about the quality of the sets, mesured in terms of few physical parameters: root mean square (RMS) deviation of the TB bands from the target bands, the resulting TB electronic gaps and effective masses.

\section{III-V Nitrides: Band Gaps and Structural Properties}
\label{sec:method}

AlN, GaN and InN nitrides are wide-gap semiconductors that usually crystallize in the wurtzite (WZ) lattice. However, under certain conditions the zincblend (ZB) crystal structure can be obtained by growing the materials on zincblende substrates. Table \ref{table:tab1} resumes their structural parameters and gaps at $\Gamma$. All these compounds, except AlN in ZB structure, are direct gap and their alloys as well their quantum wells are important from the application perspectives such as optoelectronics, because these compounds are the key constituent in blue diode lasers and LEDs \cite{vurgaftman2001} active regions.

The zincblend crystal structure consists in two inter penetrating face centered cubic (FCC) lattices, where each sub-lattice is occupied by a different chemical specie. There are 2 atoms/unit cell in a FCC structure: One is located at the origin and the other at $a(1/4,1/4,1/4)$ position, where $a$ is the lattice parameter. The wurzite unit cell has hexagonal symmetry and its unit cell has 4 basis atoms occupying the following positions: the N anions at $(a/3,2a/3,0)$ and $(2a/3,a/3,c/2)$, and the N cations at $(a/3,2a/3,3c/8)$ and $(2a/3,a/3,7c/8)$.   

\begin{table}
\scriptsize
\caption{Structural properties and gaps of the Nitride binary compounds studied in this article.}
\begin{tabular}{|l|c|c|c|c|}
\hline\hline
  Compound & $a$ (zincblend)   & $a$,$c$ (wurtzite)  & $E_{gap}$(zincblend) &  $E_{gap}$(wurtzite)  \\ \hline
  GaN      &        4.50       &     $a$ = 3.19      &      3.30            &      3.51             \\
           &                   &     $c$ = 5.18      &                      &                       \\
  AlN      &        4.38       &     $a$ = 3.11      &      4.90            &      6.23             \\
           &                   &     $c$ = 4.98      &                      &                       \\
  InN      &        4.98       &     $a$ = 3.54      &      1.94            &      1.99             \\
           &                   &     $c$ = 5.70      &                      &                       \\
 \hline\hline
\end{tabular}
\label{table:tab1}
\normalfont
\end{table}

\section{Formalism}

\subsection{Orthogonal Tight-Binding Formalism}

The description of the electronic structure within the OTB is given in the seminal paper by Slater and Koster \cite{slater1954}. Within the two center approximation, it is assumed the basis orbital forms an orthogonal set, and the Hamiltonian matrix elements between two basis orbitals are expressed only in terms of the orbital's symmetry and the distance among them, not considering the contributions from atoms localized in different lattice sites. The inclusion of $d$ orbitals in the basis is needed for a good description of the bands, but increases the computational effort. On the other hand, a device first introduced by Vogl \cite{vogl} consists in replacing the 5 $d$ orbitals for an effective excited $s$ orbital, $s^\ast$. Thus, it is employed a $sp^{3}s^{\ast}$ basis for the TB description of the electronic structure in the orthogonal formulation, avoiding this way the use of $d$ orbitals.

Within the $sp^{3}s^{\ast}$ basis, the TB bulk Hamiltonian is written as
\begin{equation}
H=\sum\limits_{ij\mu\nu}h_{ij}^{\mu\nu}c_{i\mu}^{\dagger}c_{j\nu}
\label{hamotb}
\end{equation}
where $i$ and $j$ denote the sites in the zincblende structure and $\mu$ and $\nu$ denote the atomic orbitals. The spin-orbit corrections are neglected in the parameter calculations. The $h_{ij}^{\mu\nu}$ values in \ref{hamotb} correspond to all the on-site orbital energies ($i=j$) and hoppings ($i\neq j$). In this article, the hoppings $h_{ij}^{\mu\nu}$ for AlN, GaN and InN compounds are restricted to pairs $(i,\, j)$ up to second neighbors, yielding in two kind of TB sets. Within this approach, the band structure is calculated by diagonalizing the $10\times 10$ Hamiltonian built in the basis of Bloch sums of the corresponding atomic valence orbitals.

\subsection{The Extended H\"uckel Theory}
\label{subsec:eht}

The EHT calculations shares with other Tight-Binding approaches the use of atomic orbitals as basis sets, but in comparison with the OTB formulation, the method works with explicit analytical expressions for the basis orbitals. As a result, a price to be paid is the additional calculation of the overlap matrix $\mathbf{S}$ among the basis orbitals. A common choice is to express the basis orbitals $\lbrace\Phi_{\nu}\rbrace$ as a sum of two Slater-Type Orbitals (STO) (double zeta basis). The matrix elements of the Hamiltonian in the EHT in terms of the atomic basis set are:
\begin{eqnarray}
 H_{\mu\mu} &=& \langle\Phi_{\nu} \vert H \vert \Phi_{\nu}\rangle = E_{\mu\mu}  \nonumber \\
 H_{\mu\nu} &=& \frac{1}{2}K_{EHT}\left(H_{\mu\mu} +H_{\nu\nu} \right)S_{\mu\nu} \nonumber \\
 S_{\mu\nu} &=& \langle\Phi_{\mu} \vert \Phi_{\nu}\rangle = \int{\phi_{\mu}^{\ast}\phi_{\nu}d^{3}\mathbf{r}},
\label{huckel}
\end{eqnarray}
where $K_{EHT}$ is an additional fitting parameter whose value is commonly set to 1.75 for molecules and 2.3 for solids \cite{jcerda2000}, and $S_{\mu \nu}$ is the overlap between the $\vert \Phi_{\mu}\rangle$ and $\vert \Phi_{\nu}\rangle$ orbitals. In order to perform calculations within the EHT, it is necessary to specify, for each atom type, the onsite energies ($E_s$, $E_p$ and $E_d$), the zetas of the Slater Orbitals, and the first expansion coefficient $c_1$. The second coefficient value is constrained in order to guarantee the orbital normalization. The Tight-Binding band structure is obtained by solving the generalized eigenvalue problem:

\begin{equation}
\mathbf{H}(\mathbf{k})\Psi_{i}(\mathbf{k})=E_{i}(\mathbf{k})\mathbf{S}(\mathbf{k})\Psi_{i}(\mathbf{k}),
\label{gevp}
\end{equation}
where $\Psi_{i}(\mathbf{k})$ denotes the eigenvector of the $i$th band, and $\mathbf{k}$ is the Bloch wave vector within the first Brillouin Zone. The overlap and Hamiltonian matrices, $S(\mathbf{k})$ and $H(\mathbf{k})$, are calculated through
\begin{eqnarray}
 H_{i,j}(\mathbf{k}) &=& \sum_{j',m'} e^{i\mathbf{k}\cdot\left(R_{i0}-R_{j'm'} \right)}H_{i0,j'm'}  \label{hhm} \\
 S_{i,j}(\mathbf{k}) &=& \sum_{j',m'} e^{i\mathbf{k}\cdot\left(R_{i0}-R_{j'm'} \right)}S_{i0,j'm'} \label{hsm},
\end{eqnarray}
where $i$ and $j$ label the atoms within the unit cell and $m'$ is the unit cell index. The summation indices in equations \ref{hhm} and \ref{hsm} run over all atoms $j'$ in the unit cell $m'$ which are equivalent to atom $j$ in the reference unit cell $m=0$. The real-space matrix elements $H_{i0,j'm'}$ and $S_{i0,j'm'}$ constructed between an atom $i$ in the reference  unit cell and atom $j'$ in cell $m'$ are calculated through the Extended H\"uckel prescription, Eqs. (\ref{huckel}).In addition, hoppings were restricted to sites with inter-atomic distances less than 9 $\textnormal{\AA}$ (cutoff radius).

\section{Simulation Annealing Procedure}  

As published in \cite{asmssc}, all TB parameters were calculated using a simulated annealing (SA) approach within the proposal of Vanderbilt and Louie \cite{vanderbilt1984}. In a few words, for a given initial set of H\"uckel parameters, their values are varied in successive Monte Carlo cycles with decreasing temperatures in order to reduce the value of the objective function $y$, namely the root mean square (RMS) deviation of the H\"uckel bands $E_{i}^{H}(\mathbf{k})$ in relation to a target band structure $E_{i}^{T}(\mathbf{k})$:
\begin{equation}
y = \sqrt{\frac{1}{nb\times nk}\sum_{i=1}^{nb}\sum_{j=1}^{nk}\left[E_{i}^{H}(\mathbf{k}_{j}) - E_{i}^{T}(\mathbf{k}_{j}) \right]^2},
\label{objfunc}
\end{equation}
where $nb$ and $nk$ denote, respectively, the number of bands and k-points. The target band structure were calculated by the {\it ab-initio} Density Functional Theory (DFT) formalism as implemented in the Abinit package \cite{gonze2009}, with plane wave cutoff energy of 40 Ha and Hartwigsen-Goedecker-Hutter pseudopotentials~\cite{krack2005}. The calculations were carried out using the Generalized Gradient Approximation (GGA) as parameterization of the exchange-correlation potential and the bands were generated along the $\Gamma-X-L-\Gamma$ lines for the zincblend structure, and $M-L-A-\Gamma-K-H$ for the wurtzite. 

As DFT underestimates the band gap value, its conduction bands were shifted by the difference between the experimental and the DFT gap values. In the minimization procedure all valence bands and the first conduction bands were included in Eq. \ref{objfunc}. An acceptable set is generated when $y\le 0.15$ eV. This approach was successful for generating acceptable and highly transferable TB sets for few III-V semiconductors compounds (AlAs, GaAs, InAs and GaP) \cite{asmssc}, as well and group-IV planar structures such as graphene, silicene and germanene \cite{asmjpcm}.

\section{Results and Discussion}
\label{sec:results} 

\subsection {OTB parameters}
\label{subsec:otbp}

As two orthogonal TB sets are presented for each compound, one set has only first-neighbor hoppings and the other set has hoppings up to second neighbors, let's denote them, respectively, as OTB $1nn$ and OTB $2nn$. For a given zincblend compound, the TB parameters in the OTB formulation can be divided in two groups: one referring to the anion atom ($a$) and other to the cation ($c$). For the $sp^{3}s^{\ast}$ basis set, the OTB parameters correspond to the onsite energies values for the anions $(E_{sa}, E_{pa}, E_{sta})$ and for the cations $(E_{sc}, E_{pc}, E_{stc})$ and the first neighbor hoppings $(V_{ss},V_{xx},V_{xy},V_{sapc},V_{pasc},V_{stapc},V_{pastc},V_{sst},V_{stst})$ and the corresponding second neighbor hoppings. In order to reduce the number of parameters to be fitted, for all OTB $1nn$ sets it is assumed $V_{sst}=0$ and $V_{ssta}=V_{sstc}=V_{ststa}=V_{ststc}=0$.

\begin{table}
	\caption{OTB sets for both $1nn$ and $2nn$ models. The top of the valence band for all compounds was set to zero.}
  \begin{tabular}{|l|r|r|r|r|r|r|}
  \hline\hline
\multicolumn{1}{|c|}{} & \multicolumn{2}{c|}{AlN} & \multicolumn{2}{c|}{GaN} & \multicolumn{2}{c|}{InN} \\  \hline
            &  $1nn$  &   $2nn$  &   $1nn$    &   $2nn$  &  $1nn$   &   $2nn$ \\ \hline
$E_{sa}$      & -4.1253 & -4.7340  & -2.9928  & -6.9892  &-12.0165  &  -3.2307 \\
$E_{pa}$      &  0.2013 & -0.2575  &  8.2421  &  0.0087  &  5.2396  &   2.0365 \\
$E_{sta}$     & 21.1872 & 26.3273  &  7.4040  & 28.5516  &  7.5473  &  25.1512 \\
$E_{sc}$      & -3.5695 &  0.3229  & -9.1457  & -1.1585  & -0.0208  &  -2.1405 \\
$E_{pc}$      & 20.3600 & 11.2192  & 18.5207  & 10.0413  & 15.4685  &   7.8761 \\
$E_{stc}$     & 14.7917 & 20.3649  & 24.7712  & 18.8317  & 14.4429  &  22.3537 \\
$V_{ss}$      &-10.6435 & -9.4920  & -9.0571  & -8.4367  & -5.5534  &  -8.0127 \\
$V_{xx}$      &  2.2916 &  3.2124  & 12.2267  &  3.2616  &  8.9166  &   6.1282 \\
$V_{xy}$      &  6.2680 &  4.7497  & 15.0712  &  5.9180  & 11.3494  &   6.4427 \\
$V_{sapc}$    &  5.2908 &  3.9682  &  8.9932  &  1.7631  &  3.4244  &   4.9504 \\
$V_{pasc}$    & 10.8227 & 12.1895  &  8.8691  & 12.0073  &  7.3719  &  12.0874 \\
$V_{stapc}$   & 16.4151 &  7.4956  &  6.6838  &  6.7579  &  4.9810  &   8.1301 \\
$V_{pastc}$   &  0.4568 &  0.0001  &  0.4338  &  0.0002  &  3.2377  &   0.0002 \\
$V_{stst}$    &  0.2000 &  0.0000  &  0.1313  &  0.0000  &  0.0216  &   0.0000 \\
$V_{sasa}$    &         & -1.4466  &          & -1.2606  &          &  -1.6509 \\
$V_{sxa110}$  &         &  0.7521  &          &  0.9329  &          &   1.0310 \\
$V_{stxa110}$ &         & -0.0144  &          & -0.0793  &          &  -0.0001 \\
$V_{xxa110}$  &         &  0.4155  &          &  0.3533  &          &   0.5828 \\
$V_{xxa011}$  &         & -0.0089  &          & -0.0020  &          &   0.0355 \\
$V_{xya110}$  &         &  1.0982  &          &  0.9027  &          &   1.2079 \\
$V_{scsc}$    &         & -0.1839  &          & -0.1811  &          &  -0.7139 \\
$V_{sxc110}$  &         &  1.1282  &          &  0.9257  &          &   0.8519 \\
$V_{stxc110}$ &         & -0.0071  &          & -0.0157  &          &  -0.0170 \\
$V_{xxc110}$  &         &  2.7809  &          &  1.9169  &          &   1.6969 \\
$V_{xxc011}$  &         & -0.1486  &          & -0.0858  &          &  -0.0263 \\
$V_{xyc110}$  &         &  0.3100  &          &  0.6717  &          &   1.3816 \\
	\hline\hline
	\end{tabular}
	\label{table:tab2}
	\normalfont
\end{table}

\subsection {H\"uckel Parameters}
\label{subsec:hpar}

 The optimized set of parameters for GaN, AlN and InN compounds are summarized in Tables \ref{table:tab3} and \ref{table:tab4}. In the SA procedure, all valence bands and the first conduction band were included, resulting a total of 5 bands for zincblend and 9 for wuztzite structures. The systematic adopted in SA procedure was first optimize the AlN, and for GaN and InN, the values of $\zeta$ and $c_1$ of N were set to the calculated values for AlN, emphasizing only orbital energies for this specie are varied.

\begin{table}
	\scriptsize
	\caption{\scriptsize Optimized parameters of the atomic orbitals (AO) basis set calculated by SA. Although all AO's are of the double-$\zeta$ Slater type, the values of the $c_2$ coefficient is not included whenever $\zeta_{2} = 25$ and, for these cases, $c_2=\sqrt{1-c_{1}^2}$. The $K$ value in Eq. \ref{huckel} was set to 2.3 and the Fermi Level was fixed to -13 eV.}
	\begin{tabular}{|l|c|c|c|c|c|}
		\hline\hline
		& AO & $\zeta_{1}$ & $c_1$ & $\zeta_{2}$ & $c_2$ \\ \hline
		\multirow{5}{*}{N}  & $2s$ & 2.4161 & 0.9399 &  & \\ & $2p$ & 1.8569 & 0.9221 & 3.4019 & 0.3870 \\ & $3d$ & 0.7243 & 0.4181 &  &  \\ \hline
		\multirow{5}{*}{Al} & $3s$ & 1.5943 & 0.6926 &  & \\ & $3p$ & 1.1362 & 0.6200 & 4.5546 & 0.7846 \\ & $3d$ & 0.7423 & 0.6570 & 4.6871 & 0.7539 \\ \hline
		\multirow{5}{*}{Ga} & $4s$ & 2.0033 & 0.6356 &  & \\ & $4p$ & 1.6068 & 0.7761 & 8.9752 & 0.6307 \\ & $4d$ & 1.0399 & 0.6364 &  &  \\ \hline
		\multirow{5}{*}{In} & $5s$ & 2.3211 & 0.7212 &  & \\ & $5p$ & 1.9476 & 0.7325 & 8.4423 & 0.6808 \\ & $5d$ & 1.3491 & 0.6423 & & \\
		\hline\hline
	\end{tabular}
	\label{table:tab3}
	\normalfont
\end{table}

As the parameters for GaN were already calculated in \cite{asmssc}, in this paper the parameters were recalculated. In Table \ref{table:tab4}, the onsite energies are given for both zincblend and wurtzite structures. The parameters were generated first for the zincblend structure; moreover, in order to test their transferability, only the onsite energies were varied in the parameterization procedure for the wurtzite structure. As previously published in \cite{asmssc} for other III-V compounds, the atomic orbital related parameters depend only on the atomic specie, being the same for both zincblend and wurtzite. For all compounds, the final RMS value was always less than 0.16, with the worst value obtained for InN. The onsite energies follow the same trend: as lighter the element, more close are the values for the wurtzite onsite energies with respect to zincblend. 

\begin{table}
	\scriptsize
	\caption{\scriptsize Optimized atomic basis orbitals on-site energies for both zincblend and wurtzite crystalline structures.}
	\begin{tabular}{|l|r|r|r|r|r|r|r|}
	\hline\hline
	\multicolumn{2}{|c|}{}  &  \multicolumn{3}{c|}{Zincblend} & \multicolumn{3}{c|}{Wurtzite} \\  \hline
	Compound &  Element &    $E_s$    &    $E_p$   &   $E_d$    &   $E_s$    &    $E_p$   &   $E_d$    \\ \hline
	AlN      &     Al   &  -12.9281  &  -8.6090  &  -4.4930  & -12.7164  &  -8.0662  &  -4.8278  \\
	&     N    & -23.5307 &  -13.4971 &  -2.5044  & -23.7799  &  -13.6877 &  -2.1462  \\
	GaN      &     Ga   &  -15.6698  &  -8.7221  &  -3.8077  & -15.5754  &  -9.1129  &  -3.4400  \\
	&     N    &  -23.8265  &  -13.4739 &  -3.0276  & -23.6683  &  -13.4821 &  -3.7490  \\
	InN      &     In   &  -15.0021  &  -8.7243  &  -1.5326  & -14.9593  &  -8.8241  &  -1.4675  \\
	&     N    &  -23.7727  &  -13.0180 &  -1.0627  & -23.7213  &  -13.0377 &  -0.8192  \\
	\hline\hline
	\end{tabular}
	\label{table:tab4}
	\normalfont
\end{table}

From Table \ref{table:tab4}, it is possible to assess the parameters transferability and even the small impact of the crystalline environment on the onsite energies. For the AlN and GaN, better RMS values were obtained and even for the InN, it is possible to assert the quality of the parameters by the following way: the wurtzite unit cell has 4 atoms, and for this structures the minimization procedure was carried out just considering a number of valence bands twice compared to the zincblend structures, whose unit cell has 2 atoms.

\subsection{Discussion}

\begin{figure}[ht]
	\begin{center}
		\includegraphics[width=8cm,height=6cm]{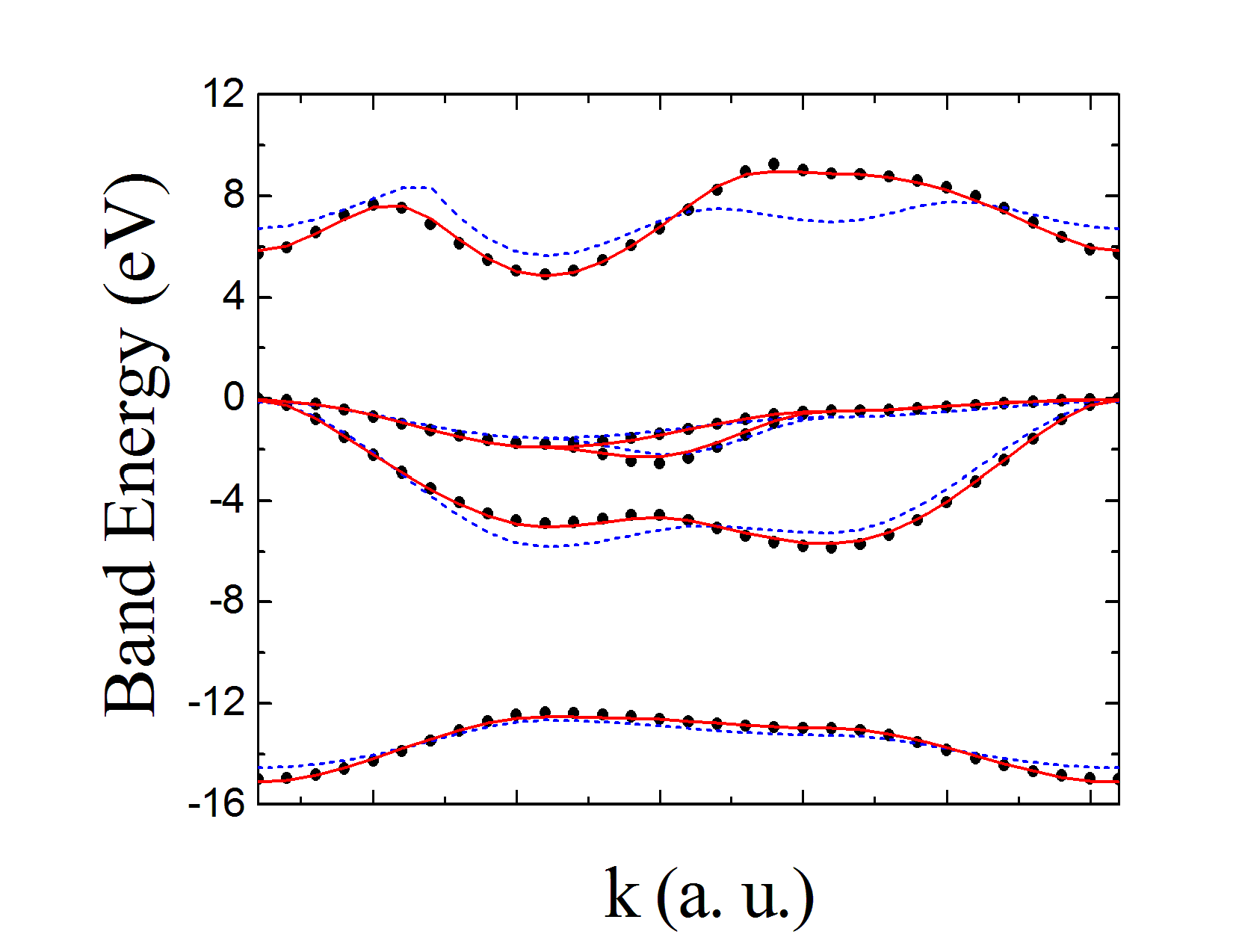}
		\caption{Comparison of the resulting bands for the $1nn$ (blue dashed) and $2nn$ TB sets with respect to the target DFT bands (dots) \label{fig1}}
	\end{center}
\end{figure}

In Table \ref{table:tabrms}, there is a summary of the final RMS values for all calculated sets by the SA method. As expected, the best parametrizations correspond to the OTB $2nn$ and the one given by the H\"uckel model. Fig. \ref{fig1} shows the calculated band structures of AlN for the $1nn$ and $2nn$ OTB models compared to the target DFT bands. AlN was choose because its the $1nn$ model result the worst RMS value of Table \ref{table:tabrms}, whereas the other models yield excellent fits. As the value of the final RMS for the H\"uckel set is very close to the $2nn$ model, the corresponding bands were not included in the Figure \ref{fig1} in order not to overload it. Notice that in the Figure the excellent overall agreement of the $2nn$ model with the DFT bands and an acceptable for the valence bands of the $1nn$ model; furthermore, even the $1nn$ model gives a bad description of the conduction band, the description of the region around the minimum is not so bad.

\begin{table}
\caption{Final values of the RMS for all TB models.}
\begin{tabular}{|l|c|c|c|c|c|c|}
	\hline\hline
	\multicolumn{1}{|c|}{}  &  \multicolumn{3}{c|}{Zincblend} & \multicolumn{1}{c|}{Wurtzite} \\  \hline
	Compound &    OTB $1nn$    &   OTB $2nn$   &   H\"uckel    &   H\"uckel  \\ \hline
	AlN      &  $0.540$  &  $0.096$  &  $0.096$   & $0.099$    \\
	GaN      &  $0.312$  &  $0.071$  &  $0.129$   & $0.068$    \\
	InN      &  $0.229$  &  $0.095$  &  $0.135$   & $0.102$    \\
	\hline\hline
\end{tabular}
\label{table:tabrms}
\normalfont
\end{table}

Aiming to quantify the quality of all TB sets, Tables \ref{table:tabgap} and \ref{table:tabmass} present, respectively, the calculated gaps and the effective mass of all compounds in the zincblend structure. Regarding the gaps, the calculated values for all TB models show deviations less than 5\% from the corresponding target values. For the effective masses, the deviations for the best models were less than 45\% and this result is acceptable, because the fit procedure adopted in this work do not use the effective masses as target values, as the TB parameters fitted by Klimeck and collaborators \cite{klim01,klim02} for some group IV and III-V semiconductors.

As expected, the OTB $2nn$ and the H\"uckel model result in a better description of the effective masses, although the $1nn$ model does gives non pathological values for the $m_{e}(\Gamma)$ and $m_{t}(X)$ masses. However, for all models, the $m_{l}(X)$ for InN is very high compared to the target value. In this case, there is an interesting feature of the present proposal: the effective masses values reflect the quality of the target band structure, and as better were the target bands better will be the values of the calculated effective masses. The so called band gap problem in the DFT calculations was overcome by adding the difference between the experimental and the DFT gap values; however, for the effective masses, there is no similar device because the values of the effective masses will depend on the correct description of the curvature bands around the edge. In the present case, the target bands were calculated within the standard DFT calculations employing the GGA approximation for the exchange-correlation potential. However, a even more precise calculation can be done within the state-of-art quasiparticle calculation based on the GW approximation \cite{rubio}

\begin{table}
	\caption{Calculated gaps in eV for all models considered.}
	\begin{tabular}{|l|c|c|c|c|c|c|}
		\hline\hline
		\multicolumn{1}{|c|}{}  &  \multicolumn{4}{c|}{Zincblend}  & \multicolumn{2}{c|}{Wurtzite} \\  \hline
		Compound &   Target  & OTB $1nn$ &  OTB $2nn$ &   H\"uckel & Target & H\"uckel  \\ \hline
		AlN      &  $4.90$   &  $5.74$   &  $4.88$    &  $4.86$    & $6.23$ & $6.21$    \\
		GaN      &  $3.30$   &  $3.38$   &  $3.40$    &  $3.12$    & $3.51$ & $3.43$    \\
		InN      &  $1.94$   &  $2.18$   &  $2.22$    &  $2.07$    & $1.99$ & $2.11$    \\
		\hline\hline
	\end{tabular}
	\label{table:tabgap}
	\normalfont
\end{table}

There are few TB parameters published in the literature for the compounds considered here. In the work of G\"urel and collaborators  \cite{gurel}, the authors use $sp^3s^{\ast}$ basis set with orbital interaction up to second neighbors, but only the hoppings between second neighbors $p$ orbitals are considered, all the others being set to zero. However, the authors in the paper not present a comparison between the resulting bands with bands calculated with other methods: just the energies in the high-symmetry points of the BZ are presented. On the other hand, in the paper of Jancu \textit{et. al.} \cite{jancu}, a $sp^3d^5s^{\ast}$ basis set is employed, with only nearest-neighbor orbital interaction. This article presents TB sets for both zincblend and wurtzite phases and the exponents of the Harrison scaling Law for the hoppings, needed for correct them in the case of deviation of the atomic position from the ideal crystal values. Thus, despite the use of $d$ orbitals, the calculated bands and the effective masses are well described in this model, being this paper a reference for TB sets for the AlN, GaN and InN compounds.

Concerning the H\"uckel parameters, this article publish the first parametrization for the AlN and InN compounds in both zincblend and wurtzite crystalline structures. Different from the OTB sets, which need the exponents of the Harrison scaling Law for correcting the hoppings when the system suffers structural deformations, the H\"uckel parameters are highly transferable. Moreover, being the hoppings proportional to the overlap between the involved orbitals, their values are corrected just recalculating the overlap and this is a great advantage of the EHT over OTB.

\begin{table}
	\caption{Calculated effective masses for all models and compounds (zincblend)}
	\begin{tabular}{|l|c|c|c|c|c|c|c|}
		\hline\hline
	\multicolumn{1}{|c}{} & \multicolumn{1}{c|}{}  &  \multicolumn{4}{c|}{Zincblend}   \\  \hline
    &  Compound &   Target  & OTB $1nn$ &  OTB $2nn$ &   H\"uckel  \\ \hline
\multirow{3}{*}{$m_{e}(\Gamma)$}		&	AlN      &  0.25   &  0.742   &  0.378    &  0.369    \\	  &	GaN      &  0.15   & 0.188 &  0.215     &  0.192    \\
	&	InN      &  0.12   &  0.133   &  0.140    &  0.152    \\
	\hline
\multirow{3}{*}{$m_{l}(X)$}		&	AlN      &  0.53   &  0.463   &  0.451    &  0.617    \\	&	GaN      &  0.5   &  2.68   & 0.546    &  0.694   \\
	&	InN      &  0.48   &  2.735   &  1.643    &  1.416   \\
	\hline 
\multirow{3}{*}{$m_{t}(X)$}		&	AlN      &  0.31   &  0.467   &  0.388    &  0.312   \\	&	GaN      &  0.3   & 0.342  &  0.296    &  0.264   \\
	&	InN      &  0.27   &  0.308   &  0.288    & 0.304   \\
	\hline\hline
	\end{tabular}
	\label{table:tabmass}
	\normalfont
\end{table}

\section{Conclusions}
\label{sec:conclusion}

The article presents a Simulated Annealing approach for calculation of the TB parameters for the group III nitrides AlN, GaN and InN. The sets are divided in two ``flavors": orthogonal basis set (OTB) and H\"uckel non-orthogonal parameters. For the former, both nearest and second nearest neighbor sets are presented for a $sp^3s^{\ast}$ basis and zincblend structure and, for the later, the H\"uckel parameters were calculated in for a $spd$ basis and for both zincblend and wurtzite phases.

\section{Acknowledgments}

The author is grateful for the financial support of the Brazilian funding agency Funda\c c\~ao de Amparo 
\`a Pesquisa do Estado do Rio de Janeiro (FAPERJ) through grants E-26/112.554/2012 and E-26/110.318/2014. The author also would like to thanks to I. A. Ribeiro for the kind revision of the manuscript.

\end{document}